\documentclass[runningheads]{llncs}
\usepackage{graphicx}
\usepackage{cite}
\usepackage{amsmath,amssymb,amsfonts}
\usepackage{float}
\usepackage{caption}
\usepackage{algorithmic}
\usepackage{makecell}
\usepackage{footnote}
\makesavenoteenv{tabular}
\makesavenoteenv{table}
\usepackage{subfigure}
\usepackage{algorithm}
\usepackage{algorithmic}
\usepackage{amsmath}
\usepackage{comment}
\usepackage{multirow}
\usepackage{xcolor}
\usepackage{eso-pic}

\AddToShipoutPictureBG{%
  \AtPageUpperLeft{%
    \makebox[\pdfpagewidth]{\raisebox{\dimexpr-\height-20pt}{%
      \small Accepted at MICCAI 2022 Workshop on Distributed, Collaborative and Federated Learning
    }}%
  }%
}

\begin{document}

\title{Cluster Based Secure Multi-Party Computation in Federated Learning for Histopathology Images}
%
%
\author{S. Maryam Hosseini\inst{1} \and
Milad Sikaroudi\inst{1}\\ \and
Morteza Babaei\inst{1,2} \and
H.R. Tizhoosh\inst{1,2,3}}
\authorrunning{S. Maryam Hosseini et al.}
%
\institute{Kimia lab, University of Waterloo, Waterloo, ON N2L 3G1, Canada \\
\email{\{sm24hoss,milad.sikaroudi\, morteza.babaie\}@uwaterloo.ca}
\and
Vector Institute, MaRS Centre, Toronto, Canada \and
Dept. of Artificial Intelligence and
Informatics, Mayo Clinic, Rochester, MN, USA \\
\email{tizhoosh.hamid@mayo.edu}}
\maketitle              
\begin{abstract}
Federated learning (FL) is a decentralized method enabling hospitals to collaboratively learn a model without sharing private patient data for training. In FL, participant hospitals periodically exchange training results rather than training samples with a central server. However, having access to model parameters or gradients can expose private training data samples. To address this challenge, we adopt secure multiparty computation (SMC) to establish a privacy-preserving federated learning framework. In our proposed method, the hospitals are divided into clusters. After local training, each hospital splits its model weights among other hospitals in the same cluster such that no single hospital can retrieve other hospitals' weights on its own. Then, all hospitals sum up the received weights, sending the results to the central server. Finally, the central server aggregates the results, retrieving the average of models' weights and updating the model without having access to individual hospitals' weights. We conduct experiments on a publicly available repository, The Cancer Genome Atlas  (TCGA). We compare the performance of the proposed framework with differential privacy and federated averaging as the baseline. The results reveal that compared to differential privacy, our framework can achieve higher accuracy with no privacy leakage risk at a cost of higher communication overhead.   

\end{abstract}
\section{Introduction}
Machine learning methods rely on a large number of data collected in a centralized location for training purposes. However, most data owners such as medical centers are not willing to share their private data with others because of privacy regulations~\cite{rieke2020future}. To address the data privacy concern, decentralized methods such as Federated learning (FL) are emerging. FL enables learning a model while all participants keep data private, sharing training results with the central server. However, authors in~\cite{zhu2019deep} have shown that sharing the model's parameters or gradients is not safe. They demonstrate that having access to the model's weight or gradients can expose training samples. Therefore, privacy-preserving methods in FL have recently been introduced to protect training samples from leakage. 
There are three different strategies for privacy-preserving FL in the literature to securely share the training results~\cite{yin2021comprehensive, kaissis2020secure}.  
\begin{itemize}
    \item \textbf{Differential Privacy (DF)}~\cite{abadi2016deep} protects privacy by adding noise to the training results before sharing with the central server. Although perturbing the training results improves the privacy of the training samples, it adversely impacts accuracy. 
    \item \textbf{Secure Multiparty Computation (SMC)}~\cite{lindell2020secure} is a privacy-preserving method, enabling hospitals to jointly compute a function on their model's weight without revealing the actual weights values. Although SMC does not perturb the training results, it has communication overhead since hospitals communicate with each other to compute the average weights.    
    \item \textbf{Homomorphic Encryption (HE)}~\cite{brutzkus2019low} relies on encoding/decoding gradients and uses encrypted data for training. 
    It allows computation on encrypted gradients and decryption of the results is equivalent to performing the same operations on gradients without any encryption. This method is efficient in terms of communication cost, however, it is computationally expensive. 
\end{itemize}

The effectiveness of DP in decentralized learning has been investigated in the healthcare domain~\cite{li2020multi, adnan2022federated}.
Authors in~\cite{li2020multi} preserve accuracy by adding Gaussian noise to the trained model weights, providing extensive experiments on MRI images. In~\cite{adnan2022federated}, the authors conduct the feasibility study of DP in federated learning. Also, the impact of the design factors of DP in decentralized learning has been investigated on histopathology images. 

SMC has played a successful role in cloud computing and the Internet of Things (IOT)~\cite{zhao2019secure}. Recently, SMC has been adopted as a privacy-preserving method in federated learning. For example, 
authors in~\cite{li2020privacy} applied chained SMC in FL to protect model weights from disclosure. In their framework, first, the central server sends one of the participants a random number. Then participants sequentially communicate with each other to compute the average of the local models. This framework imposes extreme latency and cannot be scaled since all the participant has to communicate sequentially. However, in our proposed method, communications happen in parallel within clusters. 
In this paper, we address the privacy challenges of federated learning by introducing a novel framework based on SMC. Unlike DP, SMC does not compromise the model accuracy since it does not perturb training results. In our proposed method, we divide the hospitals into small clusters.
Hospitals within each cluster collaborate to learn the summation of the local weights without having access to individual hospitals' trained weights. We perform experiments on the histopathology lung cancer dataset, comparing the performance of the proposed method with DP and baseline. 

\section{Method}

\begin{figure}
    \centering
    \includegraphics[width=0.95\textwidth]{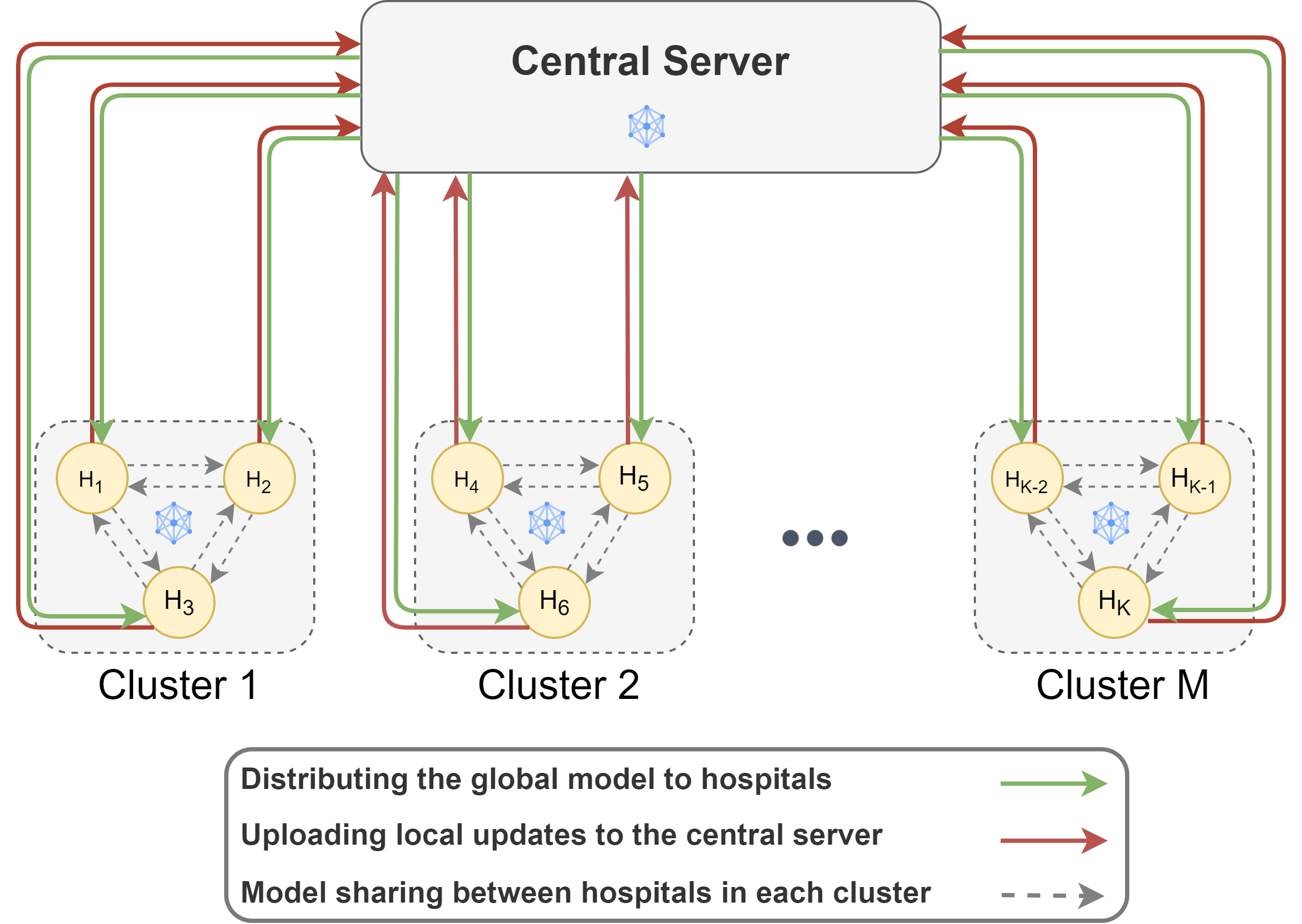}
    \caption{Cluster-based secure multi-party computation for  federated learning.}
    \label{fig:smc}
\end{figure}

\begin{algorithm}[b!]
\caption{Proposed method. There are $K$ hospitals, $M$ clusters, $T$ is the number of epochs, $E$ is the number of local epochs, $\eta$ is learning rate, $n_c$ index of all hospitals in cluster $c$.}\label{alg:fed2}
\textbf{Input}: $M, C, T,  w^0, \eta, n_c $\\
\textbf{Output}: $w^{T-1}$
\begin{algorithmic}[1] 
\FOR{$t=0, \dots, T-1$}
\STATE Server sends $w^t$ to all hospitals \\
\% \textit{Step1: Local Training}
\FOR{$k=1,2,\dots, K$}
\STATE 
$w_k^{t+1} \leftarrow \textbf{LocalTraining}(k,w^t, \eta)$ \text{\% update weights}
\ENDFOR \\
\% \textit{SMC}\\
$R_k^c \leftarrow 0$
\FOR{$c=1,2,\dots, M$}
\FOR{$k \in n_c$}
\FOR{$i \in n_c$}
\STATE {$R_k^c +=\beta_{i,k}^c w_i^t $}
\ENDFOR
\STATE Hospital $k$ feedbacks $R_k^c$ to the central server.
\ENDFOR
\ENDFOR\\
\% \textit{Step3: Aggregation}
\STATE Server updates $w^{t+1}$ as 
$$w^{t+1} \leftarrow \frac{1}{K} \sum_{k=1}^K R_k^c $$
\ENDFOR 
\STATE \textbf{return} $w^{T-1}$
\end{algorithmic}
\vspace{10pt}
{\textbf{LocalTraining}$(i,w_t, \eta):$}
\begin{algorithmic}[1]{
\STATE $\mathcal{B} \gets (\text{split dataset of } i \text{th hospital into batches of size $B$})$\\
\FOR{\text{local epoch $j = 1,2,\dots,E$}}
\FOR{\text{batch $b \in \mathcal{B}$}}
\STATE {$w \gets w_t - \eta \nabla F_k(w_t; b)$} $\quad$  \% $F_k(.)$ \textit{is the loss function for hospital $k$}
\ENDFOR 
\ENDFOR 
\STATE return $w$
}
\end{algorithmic}
\end{algorithm}
In this section, we introduce our proposed SMC-based FL method in detail. Figure~\ref{fig:smc} represents our proposed framework for cluster-based SMC. 
Before training, hospitals need to be divided into multiple groups. Clustering can be performed in different ways depending on three factors: the geographical distance between hospitals, hardware resources in each hospital, and network communication types deployed in each hospital.  
For instance, if hospitals are geographically far from each other, hospitals closer to each other can form a cluster. Another real-world scenario is that hospitals may indeed have different hardware resources to train the model causing latency and leading to asynchronous schedules. In these situations, one way to cluster hospitals is to group them into clusters of different sizes to improve total communication overhead between hospitals and the central server. Finally, network communication type is another important factor that impacts clustering in real-world scenarios. Different hospitals may have deployed different communication protocols and APIs to send/receive updates to/from other hospitals. We can group hospitals with the same communication protocols in the same cluster. 

In this work, we assume that all hospitals are placed geographically at the same distance from each other, have the same hardware resource and communication prototype. As such, we randomly select hospitals and form clusters of the same size. More specifically, given
$K$ hospitals, which will be equally divided into $M$ clusters with size $N=K/M$. Each hospital belongs to one cluster which is denoted by $c=\{1,\dots, M\}$. Hospital $k$ in cluster $c$ is represented by $H_k^c$. 
The set $n_c$ with length $N$ represents indexes of all hospitals in cluster $c$. 

Model training in our proposed approach is performed in three steps. 

\textbf{Step1: Local Training.} All hospitals train the model with their local data, updating the model. We denote model parameters trained by the $k$th hospital with $w_k$. 

\textbf{Step2: SMC.} Hospital $H_k^c$ generates $N$ random numbers $\{\beta^c_{k,j}| 0<\beta^c_{k,j}<1, j\in n_c \}$ that sum up to one. 
\begin{equation}
    \sum_{j\in n_c} \beta^c_{k,j} = 1
\end{equation}
Then, each hospital $k$ in cluster $c$, $H_k^c$, sends portions of its own locally trained model parameters to each of $N-1$ neighbours in cluster $c$. Mathematically, $H_k^c$ sends  $\beta_{k,j}^c w_k$ to hospital $j$ for all $j\in n_c$. 
In the end, the $k$th hospital will have some portion of its own, and some portion of its $N-1$ neighbor's model parameters as follows:
\begin{equation}
\label{H_kc}
\mathcal{H}_k^c: R_k^c = \sum_{i\in n_c}\beta_{i,k}^c w_i
\end{equation}

\textbf{Step3: Aggregation.}
Finally, each hospital sends $R_k^c$ to the central server, and the central server takes the average of $R_k^c$ of all the hospitals in all clusters as follows: 

\begin{equation}
\label{eq:server}
w =\frac{1}{K} \sum_{c=1}^{M} \sum_{k\in n_c} R_k^c= \frac{1}{K} \sum_{c=1}^{M} \sum_{k\in n_c} \sum_{i\in n_c}\beta_{i,k}^c w_i
\end{equation}

If we exchange the position of the two summations in Eq.~\ref{eq:server}, we will get
\begin{equation*}
    \begin{split}
    \label{proof}
        w &=  \frac{1}{K} \sum_{c=1}^{M} \sum_{i\in n_c} \underbrace{ \sum_{k\in n_c} \beta_{i,k}^c}_{1} w_i \\
        &=  \frac{1}{K} \sum_{c=1}^{M} \sum_{i\in n_c} w_i =  \frac{1}{K} \sum_{i=1}^{K} w_i
    \end{split}
\end{equation*}

As shown above, the central server can receive the exact average weights without having access to the weights of each individual hospital. 
These steps have been summarized in Algorithm~\ref{alg:fed2}.

\begin{table}
	\begin{minipage}{0.5\linewidth}
		\caption{The summary of the dataset~\cite{adnan2022federated}.}
		\label{table:student}
            \centering 
            \begin{tabular}{l c c} 
            \hline\hline 
            \textbf{Client} & \textbf{\# Slides} & \textbf{\# Patches} \\ [0.4ex] 
            \hline 
            C1: Int. Gen. Cons. & 267 & 66,483 \\ 
            C2: Indivumed & 211 & 52,539 \\
            C3: Asterand & 207 & 51,543 \\
            C4: Johns Hopkins & 199 & 49,551 \\
            C5: Christiana H. & 223 & 55,527 \\
            C6: Roswell Park & 110 & 27,390 \\ [1ex] 
            \hline 
            \end{tabular}
	\end{minipage}\hfill
	\begin{minipage}{0.45\linewidth}
		\centering
		\includegraphics[width=0.95\textwidth]{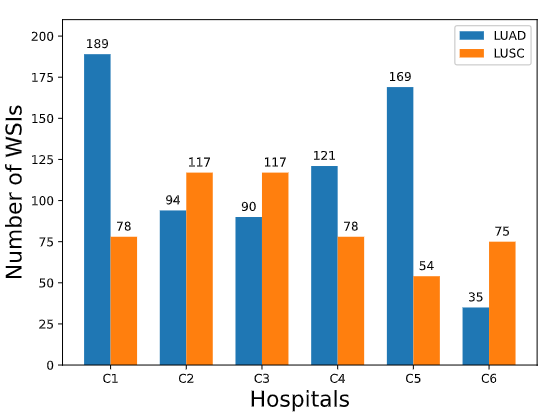}
		\captionof{figure}{Label distribution in dataset.}
		\label{fig:student}
	\end{minipage}
\end{table}

\section{Experiments and Results}

\subsection{Datasets}
We evaluate our proposed privacy-preserving FL on The Cancer Genome Atlas (TCGA)~\cite{tcga, weinstein2013cancer} dataset, the largest publicly available archive of the histopathology whole slide images (WSIs). This annotated dataset has more than $30,000$ H\&E stained WSIs that have been collected from various medical centers all over the world. 
To validate the proposed method, we select TCGA WSIs diagnosed with non-small cell lung cancer (NSCLC) to construct a dataset of multiple institutions. This cancer has two frequent subtypes, namely
    \begin{itemize}
        \item Lung Adenocarcinoma (LUAD)
        \item Lung Squamous Cell Carcinoma (LUSC).
    \end{itemize}
This study includes hospitals that have WSIs from both LUAD and LUSC subtypes. In TCGA, only six hospitals met this requirement. Therefore, we collected WSIs diagnosed with NSCLC from those six hospitals to create the dataset with six participants.   
WSIs are extremely large images of size up to $50,000 \times 50,000$ pixels. Therefore, they cannot directly be fed into any neural network. The common approach to deal with these images is to divide them into patches of smaller sizes for further analysis~\cite{hou2016patch}.
We divide the selected WSIs into patches of size $1000 \times 1000$ pixels. 
Due to space limitation, we refer readers to~\cite{adnan2022federated} for more details on patch extraction and selection of the lung dataset that has been used in our experiments. 
The statistics of this dataset for each hospital are presented in Table ~\ref{table:student} and Figure ~\ref{fig:student}. The dataset of each hospital has been randomly divided into $80\%$ and $20\%$ groups for training and testing purposes, respectively. 
\begin{figure}[hbt!]
    \centering
    \includegraphics[width=0.95\textwidth]{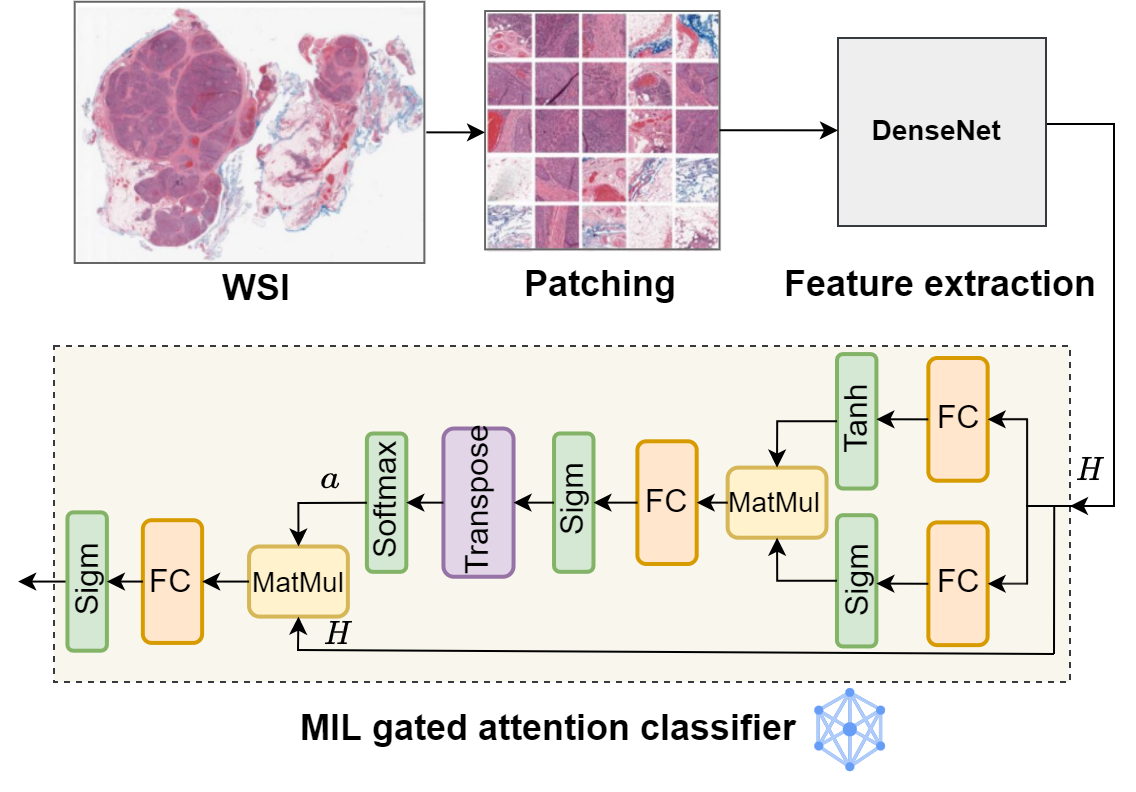}
    \caption{The illustration of the end to end training procedure. First WSIs are divided into patches of size $1000 \times 1000$. Next the features of patches are extracted using DensNet121. Finally, those features of patches are fed into MIL gated attention classifier.}
    \label{fig:network}
\end{figure}
\subsection{Experimental Details}
Figure~\ref{fig:network} illustrates WSI preprocessing as well as the model used to classify lung samples into LUAD and LUSC subtypes. 
As shown in this figure, for the classification of lung histopathology WSIs, we first employ pretrained DenseNet121~\cite{huang2017densely} to extract features of length $1024$ for each patch. Next, We employ attention-gated multiple instance learning (MIL) to combine feature vectors of patches of each WSI, creating a feature of size $1024$ for each WSI classification \cite{ilse2018attention}.
The reason why we use MIL is that when we divide a WSI into multiple patches, we are dealing with instances for which only a single WSI level label, medical diagnosis, is provided. Therefore, we require multiple instance learning (MIL) architecture to learn a model that can predict the WSI label given a bag of instances (patches). 
The attention-based MIL architecture enables the model to combine the features of patches to create one feature vector of length $1024$ that will be used for the classification of WSI. This architecture aggregates feature vectors of those patches such that key patches are assigned relatively higher weights.  
The high-level structure of the MIL classifier has been visualized in Figure~\ref{fig:network}. The MIL gated attention classifier is the network that we learn in a decentralized federated learning fashion. Due to space limitations, we refer readers to~\cite{ilse2018attention} for more detail on this MIL network.
The histopathology lung dataset includes data from six hospitals. We divide those $K=6$ hospitals into $M=2$ clusters of size $N=3$. We deploy DP according to~\cite{li2020multi} with additive Gaussian noise standard deviation of $0.03$. 
The standard deviation has been selected to have the highest possible privacy while the classification performance is still acceptable. For all these three methods, we use an Adam optimizer with the following hyper-parameter values, epochs=300, batch size=32, number of local epochs=1, and learning rate=0.009.

\subsection{Results and Discussions}
In this section, we present our experimental results on the lung histopathology dataset. We compare our proposed SMC based method with the baseline which is FedAVG~\cite{mcmahan2017communication} without any privacy-preserving consideration. We also compare our method with DP which has been implemented on top of FedAvg. An ideal privacy-preserving method has to have a closed performance to the baseline while preserving privacy of training results.  
Table~\ref{table:experiment} shows the performance of each method for each hospital in terms of accuracy and F1 Score. As represented, in each hospital, the proposed method has a closed performance to the baseline and outperforms DP. Additionally, the average performance of our method surpasses DP. 
Figure~\ref{fig:experiment1} and~\ref{fig:experiment2} compare methods in terms of the average testing accuracy and average training loss of participant hospitals for 300 rounds of training communication between hospitals and the central server. As can be seen, the proposed method performs close to the baseline, surpassing DP.     
To eliminate the impact of random parameters in our experiments, we repeated all the experiments five times and all the results have been provided by taking the average over these five realizations. 
\begin{table}
	\begin{minipage}{0.6\linewidth}
		\caption{Experimental results.}
		\label{table:experiment}
		\centering 
        \begin{tabular}{l c | c c c} 
            \hline\hline 
            \textbf{Client} & \textbf{Method} & \textbf{ACC} & \textbf{F1-Score} \\ [0.4ex] 
            \hline 
            \multirow{3}{4em}{C1} & FedAvg & 76.38 & 82.51 \\
             & DP & 66.12 & 69.89 \\
             & Proposed & 75.01 & 81.08 \\[1ex] 
            \hline 
            \multirow{3}{4em}{C2} & FedAvg & 85.46 & 87.63 \\
             & DP & 79.06 & 81.12 \\
             & Proposed & 87.20 & 89.03 \\[1ex] 
            \hline 
            \multirow{3}{4em}{C3} & FedAvg & 81.54 & 80.96 \\
             & DP & 74.40 & 70.27 \\
             & Proposed & 80.95 & 80.47 \\[1ex] 
            \hline 
            \multirow{3}{4em}{C4} & FedAvg & 75.01 & 82.74 \\
             & DP & 69.37 & 73.84 \\
             & Proposed & 75.62 & 83.12 \\[1ex]
             \hline
             \multirow{3}{4em}{C5} & FedAvg & 73.33 & 82.31 \\
             & DP & 64.87 & 68.54\\
             & Proposed & 68.88 & 78.58\\[1ex] 
            \hline 
            \multirow{3}{4em}{C6} & FedAvg & 68.18 & 66.74 \\
             & DP & 68.18 & 63.34\\
             & Proposed & 69.31 & 66.78 \\[1ex] 
            \hline 
            \hline
            \multirow{3}{4em}{Avg} & FedAvg & 76.65 & 80.48 \\
             & DP & 70.33 & 71.16 \\
             & Proposed & 76.16 & 79.84 \\[1ex] 
        \end{tabular}
	\end{minipage}\hfill
	\begin{minipage}{0.45\linewidth}
		\centering
		\includegraphics[width=0.95\textwidth]{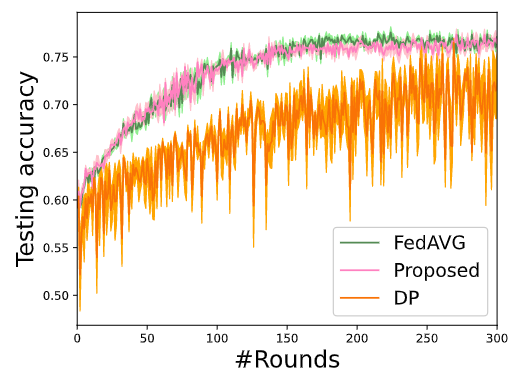}
		\captionof{figure}{The average testing accuracy for 300 rounds of training over all hospitals.}
		\label{fig:experiment1}
		\centering
		\includegraphics[width=0.95\textwidth]{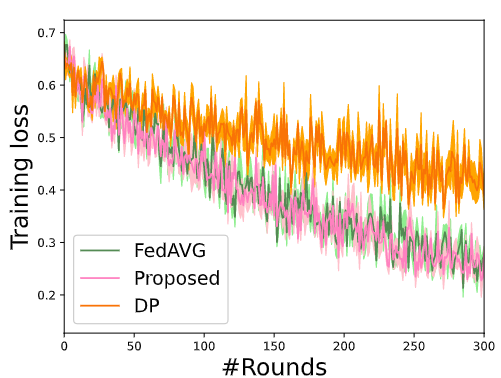}
		\captionof{figure}{The average training loss for 300 rounds of training over all hospitals.}
		\label{fig:experiment2}
	\end{minipage}
\end{table}

\section{Conclusions}
In this paper, we addressed the privacy-preserving challenge of the federated learning. We have proposed cluster-based SMC to protect individual hospitals' model parameters from disclosure. In the proposed method, neither participant hospitals nor the central server has access to model weights of individual hospitals; however, weights average can be recovered at the central server. Our experimental results suggested that the proposed method outperforms DP in terms of accuracy and F1 Score at the expense of more communication overhead. However, we believe that having slight communication overhead to get higher accuracy is most likely acceptable in the medical domain. Additionally, each hospital needs to perform preprocessing to find suitable additive noise standard deviation in DP method. However, our proposed method does not require any preprocessing since it does not have any hyper-parameter. Therefore, depending on the application, applying cluster-based SMC for privacy-preserving purposes might be preferable compared to other privacy-preserving method such as DP. 

%

%

\end{document}